\documentclass[twocolumn,aps,prb,superscriptaddress,groupedaddress,showpacs]{revtex4}
\usepackage{epsfig}
\begin{document}
\title{
\Large\bf  Field theory of bi- and tetracritical points: Relaxational dynamics}
\author{R. Folk}\email{folk@tphys.uni-linz.ac.at}
\affiliation{Institute for Theoretical Physics, Johannes Kepler
University Linz, Altenbergerstrasse 69, A-4040, Linz, Austria}
 \author{Yu. Holovatch}\email[]{hol@icmp.lviv.ua}
 \affiliation{Institute for Condensed Matter Physics, National
Academy of Sciences of Ukraine, 1~Svientsitskii Str., UA--79011
Lviv, Ukraine} \affiliation{Institute for Theoretical Physics,
Johannes Kepler University Linz, Altenbergerstrasse 69, A-4040,
Linz, Austria}
  \author{G. Moser}\email[]{guenter.moser@sbg.ac.at}
\affiliation{Department for Material Research and Physics, Paris Lodron University
Salzburg, Hellbrunnerstrasse 34, A-5020 Salzburg, Austria}
\date{\today}
\begin{abstract}
We calculate the relaxational dynamical critical behavior of systems
of $O(n_\|)\oplus O(n_\perp)$ symmetry by renormalization group
method within the minimal subtraction scheme in two loop order. The
three different bicritical static universality classes
previously found for such systems  correspond to three different
dynamical universality classes within the static borderlines. The
Heisenberg and the biconical fixed point lead to strong dynamic
scaling whereas in the region of stability of the decoupled fixed
point weak dynamic scaling holds. Due to the neighborhood of the
stability border between the strong and the weak scaling dynamic
fixed point corresponding to the static biconical and the decoupled
fixed point a very small dynamic transient exponent, of
$\omega_v^{{\cal B}}=0.0044$, is present in the dynamics for the
physically important case $n_\|=1$ and $n_\perp=2$ in $d=3$.

\end{abstract}
\maketitle

\section{Introduction}

The phase diagram of systems with $O(n_\|)\oplus O(n_\perp)$
symmetry contains several phases meeting in a multicritical point.
In Ref.\cite{partI} (henceforce called part I) it was shown that the static critical properties
can be quantitatively analyzed from field theoretic functions
in two loop order if one uses resummation.  As an example we have in
mind an antiferromagnet in an external magnetic field (with $n_\|=1$
and $n_\perp=2$), although other physical examples with different
values of order parameter (OP) components may be considered.

In order to get more insight in the dynamical critical properties
near such a multicritical point we reconsider the simplest dynamical
model possible for $O(n_\|)\oplus O(n_\perp)$ symmetric systems. In
such a dynamical model one assumes relaxational behavior for the two
OPs $\vec{\phi}_\|$ and $\vec{\phi}_\perp$. This model has been
briefly  studied\cite{dohmjanssen77} on the basis of the static one
loop results \cite{konefi76}.
Meanwhile\cite{capevi03,Prudnikov98,partI} it has been shown that
the one  loop results\cite{konefi76,Lyuksyutov75} are considerably
changed in higher loop order concerning the  regions of different
static multicriticality in the space of OP components $n_\|$ and
$n_\perp$. For integer order parameter components only a system
$n_\|=1$ and $n_\perp=2$ belongs to the universality class
characterized by the biconical fixed point (FP) indicating
tetracritical behavior - if the physical system lies in the
attraction region of the FP.\cite{note1}

The paper is organized as follows: In chapter \ref{dynmod} we define
the dynamical model, then in chapter \ref{ren} the dynamical field
theoretic functions are introduced and the results in two loop order
are presented. From these results the FP and dynamical exponents are
calculated in chapter \ref{asymp} and the stability  of the FP is
considered in chapter \ref{dynstab}. Due to the small dynamic
transient exponent found, the effective - nonasymptotic - dynamical
behavior is studied  in detail in chapter \ref{nonasymp}.
Finally conclusions and an outlook to subsequent research of
extended dynamical models is given.

\section{Dynamical model \label{dynmod}}
The results obtained in part I for the statics of systems with
$O(n_\|)\oplus O(n_\perp)$ symmetry are applied to the
critical dynamics if the system dynamics is described by two
relaxational equations for the OP components $\vec{\phi}_{\perp 0}$
and $\vec{\phi}_{\| 0}$ in the two subspaces. Correspondingly two
kinetic coefficients $\mathring{\Gamma}_\perp$ and
$\mathring{\Gamma}_\|$ have to be introduced. The model A  type
equations are
\begin{eqnarray}
\label{dphiperp}
\frac{\partial \vec{\phi}_{\perp 0}}{\partial t}&=&-\mathring{\Gamma}_\perp
\frac{\delta {\mathcal H}_{Bi}}{\delta \vec{\phi}_{\perp 0}}+\vec{\theta}_{\phi_\perp} \ , \\
\label{dphipar}
\frac{\partial \vec{\phi}_{\|0}}{\partial t}&=&-\mathring{\Gamma}_\|
\frac{\delta {\mathcal H}_{Bi}}{\delta \vec{\phi}_{\|0}}+\vec{\theta}_{\phi_\|} \, .
\end{eqnarray}
The stochastic forces $\vec{\theta}_{\phi_\perp}$ and $\vec{\theta}_{\phi_\|}$
fulfill Einstein relations
\begin{eqnarray}
\label{thetaperp}
\langle\theta_{\phi_\perp}^\alpha(x,t)\ \theta_{\phi_\perp}^\beta
(x^\prime,t^\prime)\rangle \!\!\!&=&\!\!\!
2\mathring{\Gamma}_\perp\delta(x-x^\prime)\delta(t-t^\prime)\delta^{\alpha\beta}
\ , \\
\label{thetapara}
\langle\theta_{\phi_\|}^i(x,t)\ \theta_{\phi_\|}^j(x^\prime,t^\prime)
\rangle \!\!\!&=&\!\!\! 2\mathring{\Gamma}_\|\delta(x-x^\prime)\delta(t-t^\prime)
\delta^{ij} \ ,
\end{eqnarray}
with indices $\alpha,\beta=1,\dots , n_\perp$ and $i,j=1,\dots ,n_\|$ corresponding
to the two subspaces.
The static functional ${\mathcal H}_{Bi}$ is defined as
\begin{eqnarray}\label{hbicrit}
{\mathcal H}_{Bi}\!=\!\int\! d^dx\Bigg\{\frac{1}{2}\mathring{r}_\perp\vec{\phi}_{\perp 0}
\cdot\vec{\phi}_{\perp 0}+\frac{1}{2}\sum_{\alpha=1}^{n_\perp}\nabla_\alpha\vec{\phi}_{\perp 0}\cdot
\nabla_\alpha\vec{\phi}_{\perp 0}  \nonumber \\
+\frac{1}{2}\mathring{r}_\|\vec{\phi}_{\| 0}
\cdot\vec{\phi}_{\| 0}+\frac{1}{2}\sum_{i=1}^{n_\|}\nabla_i\vec{\phi}_{\| 0}\cdot
\nabla_i\vec{\phi}_{\| 0}
+\frac{\mathring{u}_\perp}{4!}\Big(\vec{\phi}_{\perp 0}\cdot\vec{\phi}_{\perp 0}\Big)^2 \nonumber \\
+\frac{\mathring{u}_\|}{4!}\Big(\vec{\phi}_{\| 0}\cdot\vec{\phi}_{\| 0}\Big)^2
+\frac{2\mathring{u}_\times}{4!}\Big(\vec{\phi}_{\perp 0}\cdot\vec{\phi}_{\perp 0}\Big)
\Big(\vec{\phi}_{\| 0}\cdot\vec{\phi}_{\| 0}\Big)
\Bigg\} \ .
\end{eqnarray}
The properties and renormalization of the static vertex functions
following from ${\mathcal H}_{Bi}$ have already been presented in
part I. There  in resummed two-loop approximation it has been
shown that within a small region in the space of the spatial
dimension and the OP components the biconical FP is stable (e.g. for
$n_\|=1$ and $n_\perp=2$ at $d=3$).

\section{Renormalization, field theoretic functions \label{ren}}

From the dynamic equations (\ref{dphiperp}) and (\ref{dphipar}) a
functional may be derived which allows the calculation of dynamic
vertex functions in perturbation theory (for an overview see
\cite{fomorev06}). Within this dynamic functional additional
auxiliary densities $\vec{\tilde{\phi}}_{\perp 0}$ and
$\vec{\tilde{\phi}}_{\| 0}$ are introduced\cite{bauja76} . Recently
it has been shown \cite{fomo02a} that the dynamic two point
functions have a general structure, which is in the current model
\begin{eqnarray}
\label{gammappvert}
\mathring{\Gamma}_{\phi_\perp\tilde{\phi}_\perp}\big(\xi_\perp,\xi_\|,k, \omega\big)
&=&-i\omega \mathring{\Omega}_{\phi_\perp\tilde{\phi}_\perp}
\big(\xi_\perp,\xi_\|,k, \omega\big) \nonumber \\
&&+\mathring{\Gamma}^{(2,0)}_{\perp\perp}\big(\xi_\perp,\xi_\|,k\big)\mathring{\Gamma}_\perp \ ,
\end{eqnarray}
\begin{eqnarray}
\label{gammapppara}
\mathring{\Gamma}_{\phi_\|\tilde{\phi}_\|}\big(\{\xi_\perp,\xi_\|,k, \omega\big)
&=&-i\omega
\mathring{\Omega}_{\phi_\|\tilde{\phi}_\|}\big(\{\xi_\perp,\xi_\|,k, \omega\big)
\nonumber \\
&&+\mathring{\Gamma}^{(2,0)}_{\|\|}\big(\xi_\perp,\xi_\|,k\big)\mathring{\Gamma}_\| \ ,
\end{eqnarray}
where
$\mathring{\Gamma}^{(2,0)}_{\perp\perp}\big(\xi_\perp,\xi_\|,k\big)$
and $\mathring{\Gamma}^{(2,0)}_{\|\|}\big(\xi_\perp,\xi_\|,k\big)$
are the static two point vertex functions discussed in part I. The
functions
$\mathring{\Omega}_{\phi_\perp\tilde{\phi}_\perp}\big(\xi_\perp,\xi_\|,k,
\omega\big)$ and
$\mathring{\Omega}_{\phi_\|\tilde{\phi}_\|}\big(\xi_\perp,\xi_\|,k,
\omega\big)$ have to be determined within dynamic perturbation
expansion. All functions in (\ref{gammappvert}) and
(\ref{gammapppara}) depend besides the correlation functions
$\xi_\perp$, $\xi_\|$, the wave vector modulus $k$, and the
frequency $\omega$, also on the static couplings
$\mathring{u}_\perp$, $\mathring{u}_\times$ and $\mathring{u}_\|$.
The functions
$\mathring{\Omega}_{\phi_{\alpha_i}\tilde{\phi}_{\alpha_i}}$  (with
$\alpha_i=\perp,\|$) additionally depend on the two kinetic
coefficients $\mathring{\Gamma}_\perp$ and $\mathring{\Gamma}_\|$.
As we will see below, the genuine representation
(\ref{gammappvert}), (\ref{gammapppara}) that allows to single out
contributions from merely static vertex functions into dynamic ones
essentially simplifies cumbersome calculations and enables one  to
effectively proceed with calculation of the dynamic RG perturbative
expansions.

\subsection{Renormalization of the dynamic parameters}

The renormalization of the static quantities appearing in
(\ref{hbicrit}) has been presented in part I in detail  and
explicitly performed in the minimal subtraction RG
scheme\cite{Schloms} directly at $d=3$ to the two-loop order. The
resulting renormalization factors and field theoretic functions
($\zeta$- and $\beta$-functions) remain valid also in dynamics.
Additional renormalizations are necessary for dynamic quantities.
Within the current dynamic model only the auxiliary densities and
the kinetic coefficients have to be renormalized.

The renormalized counterparts of the auxiliary densities are defined
as
\begin{equation}\label{phitilderen}
\vec{\tilde{\phi}}_{\perp
0}=Z_{\tilde{\phi}_\perp}^{1/2}\vec{\tilde{\phi}}_{\perp} \ , \qquad
\vec{\tilde{\phi}}_{\|
0}=Z_{\tilde{\phi}_\|}^{1/2}\vec{\tilde{\phi}}_{\|}\, .
\end{equation}
The renormalized kinetic coefficients are introduced as
\begin{equation}\label{Gammaren}
\mathring{\Gamma}_\perp=Z_{\Gamma_\perp}\Gamma_{\perp} \ , \qquad
\mathring{\Gamma}_\|=Z_{\Gamma_\|}\Gamma_{\|} \, .
\end{equation}
Relation (\ref{phitilderen}) and the renormalization of the OP densities
$\vec{\phi}_{\perp 0}$ and $\vec{\phi}_{\| 0}$ introduced in part I imply for
the dynamic vertex functions the renormalization
\begin{eqnarray}
\label{gppperen}
\Gamma_{\phi_\perp\tilde{\phi}_\perp}&=&Z_{\phi_\perp}^{1/2}Z_{\tilde{\phi}_\perp}^{1/2}
\mathring{\Gamma}_{\phi_\perp\tilde{\phi}_\perp}
 \ , \\
\label{gppparen}
\Gamma_{\phi_\|\tilde{\phi}_\|}&=&Z_{\phi_\|}^{1/2}Z_{\tilde{\phi}_\|}^{1/2}
\mathring{\Gamma}_{\phi_\|\tilde{\phi}_\|} \ .
\end{eqnarray}
From the above relations and the structure of the dynamic two point vertex
functions presented in (\ref{gammappvert}) and (\ref{gammapppara}) follows that the
renormalization factors of the kinetic coefficients $\Gamma_\perp$ and $\Gamma_\|$
in case of the absence of mode couplings are determined by the
corresponding renormalization factors of the auxiliary densities. This leads to  the
relations
\begin{equation}\label{ZGammarel}
Z_{\Gamma_\perp}=Z_{\phi_\perp}^{1/2}Z_{\tilde{\phi}_\perp}^{-1/2} \ , \quad
Z_{\Gamma_\|}=Z_{\phi_\|}^{1/2}Z_{\tilde{\phi}_\|}^{-1/2} \, .
\end{equation}
The static renormalizaton factors $Z_{\phi_\perp}$ and $Z_{\phi_\|}$ have been
introduced in Eq. (4) of part I.

\subsection{Dynamic $\beta$- and $\zeta$-functions in two loop order}

Quite analogous to statics in part I we will use the uniform definition
\begin{equation}\label{zetafu}
\zeta_{a_i}(\{u\},\Gamma_\perp,\Gamma_\|)=\frac{d\ln Z_{a_i}^{-1}}{d\ln\kappa}
\end{equation}
for the $\zeta$-functions also in dynamics, where $a_i$ is now a
placeholder for any auxiliary density or kinetic coefficient,
$\kappa$ is the scaling parameter, and
$\{u\}=\{u_\perp,u_\times,u_\|\}$ is the set of static couplings.
From perturbation expansion the resulting two loop expressions for
the $\zeta$-functions of the kinetic coefficients $\Gamma_\perp$ and
$\Gamma_\|$ read
\begin{eqnarray}
\label{zetagperp}
\zeta_{\Gamma_\perp}&=&\frac{n_\perp+2}{36}\ u_\perp^2\left(3\ln\frac{4}{3}
-\frac{1}{2}\right)
 \\
&+&\frac{n_\|}{36}\ u_\times^2\left[\frac{2}{v}\ln\frac{2(1+v)}{2+v}
+\ln\frac{(1+v)^2}{v(2+v)}-\frac{1}{2}\right] \, , \nonumber\\
\label{zetagpara}
\zeta_{\Gamma_\|}&=&\frac{n_\|+2}{36}\ u_\|^2\left(3\ln\frac{4}{3}
-\frac{1}{2}\right)   \\
&+&\frac{n_\perp}{36}\ u_\times^2\left[2v\ln\frac{2(1+v)}{1+2v}
+\ln\frac{(1+v)^2}{1+2v}-\frac{1}{2}\right] \, . \nonumber
\end{eqnarray}
The important dynamic parameter is the time scale ratio
\begin{equation}\label{wdef}
v=\frac{\Gamma_\|}{\Gamma_\perp}
\end{equation}
between the two kinetic coefficients $\Gamma_\perp$ and $\Gamma_\|$,
which have been already introduced in (\ref{zetagperp}) and
(\ref{zetagpara}). From the above definition of the time scale ratio
and the definition of the $\zeta$-functions in (\ref{zetafu}) the
$\beta$-function of $v$ is determined by
\begin{equation}\label{betawdef}
\beta_v\equiv\kappa\frac{dv}{d\kappa}=v(\zeta_{\Gamma_\|}-\zeta_{\Gamma_\perp})
\end{equation}
where the derivative is taken at fixed unrenormalized quantities. Inserting
(\ref{zetagperp}) and (\ref{zetagpara}) into (\ref{betawdef}) the two
loop expression of the $\beta$-function of $v$ reads\cite{dohmjanssen77,dohmKFA}
\begin{eqnarray} \label{betaw}
\beta_v&=&\frac{v}{72}\Bigg\{\Big[(n_\|+2)u_\|^2-(n_\perp+2)u_\perp^2\Big](6\ln\frac{4}{3}-1) \nonumber \\
&-&n_\|\ u_\times^2\left[\frac{4}{v}\ln\frac{2(1+v)}{2+v}
+2\ln\frac{(1+v)^2}{v(2+v)}-1\right]   \\
&+&n_\perp\ u_\times^2\left[4v\ln\frac{2(1+v)}{1+2v}
+2\ln\frac{(1+v)^2}{1+2v}-1\right]\Bigg\} \, . \nonumber
\end{eqnarray}
The $\beta$-function changes its sign under interchanging the parallel and perpendicular components and replacing
the time scale ratio $v$ by $1/v$.

In the nonasymptotic region where a non universal effective critical behavior may be observed
the values of the static couplings and the time scale ratio  are described by the flow equations.
For $v$ it reads
\begin{equation}  \label{flow}
l\frac{d v}{d l}=\beta_v\big(u_\|(l),u_\perp(l),u_\times(l)\big)   \, ,
\end{equation}
whereas for the static couplings Eq. (36) of part I with the Borel
resummed static $\beta$-functions are used. The asymptotics is
reached in the limit $l\to 0$ starting in the background at $l=1$
from non universal initial values of the time scale ratio and
couplings.

\section{Fixed points and dynamical critical exponents \label{asymp}}

As usual\cite{fomorev06} the two $\zeta$-functions Eqs. (\ref{zetagperp}) and
(\ref{zetagpara})   define two dynamical critical exponents that
govern the power law increase of the autocorrelation time for the
OPs $\vec{\phi}_\|$ and $\vec{\phi}_\perp$, correspondingly
\begin{equation}
z_\perp=2+\zeta_{\Gamma_\perp}^\star \qquad\mbox{and} \qquad     z_\|=2+\zeta_{\Gamma_\|}^\star \, ,
\end{equation}
where the  stable FP values of the static and dynamic parameters have been
inserted into the $\zeta$-functions, this means
$\zeta_{\Gamma_{\alpha_i}}^\star=\zeta_{\Gamma_{\alpha_i}}(\{u^\star\},v^\star)$ .
At the strong scaling FP there is only one dynamic time scale and the two
exponents are equal whereas at the weak scaling FP they are
different and define for each component, parallel and perpendicular,
the time scale.

Depending on the FP value of the time scale ratio $v$ one may obtain
strong ($v^\star\neq 0, \infty$)  or weak  ($v^\star=0,\infty$)
dynamic scaling. The dynamical FPs are calculated (see also Eq. (12)
in  Ref.\cite{dohmjanssen77}) from setting the $\beta$-function
(\ref{betaw}) equal to zero. Inserting the stable static FP values
(see Table I in part I)  into Eq. (\ref{betaw}) one then may
calculate a dynamical 'phase diagram' in the $n_\|$-$n_\perp$-plane
quite similar to the static 'phase diagram' Fig. 1 in part I.  Let
us note here, that  one can make use of two different ways to
analyze perturbative expansions within the minimal subtraction RG
scheme. The first one is the familiar $\varepsilon$-expansion, when
the FP coordinates and asymptotic critical exponents are obtained as
series in $\varepsilon=4-d$ and then evaluated at the dimension of
interest (e.g. for $d=3$). The second one relies on treatment of the
expansions in renormalized couplings directly at fixed dimension
$d=3$.\cite{Schloms} Enhanced by resummation such a scheme allows to
treat, besides the asymptotic quantities, the non-universal
effective exponents. The latter method has been applied in part I to
perform a comprehensive analysis of non-universal static behavior.
Below we will make use of the static results obtained there to
proceed with the analysis of (asymptotic and effective) dynamical
critical behavior.

To summarize an outcome of the static FP stability
analysis,\cite{konefi76,Lyuksyutov75,Prudnikov98,partI,capevi03} let
us recall that, depending on the $n_\|$, $n_\perp$ values the
critical behavior is governed by one of the three non-trivial FPs:
(i) isotropic Heisenberg FP ${\mathcal H}(n_\perp+n_\|)$ with
$u_\|^\star=u_\perp^\star=u_\times^\star=u^\star$; (ii) decoupling
FP ${\mathcal D}$  with $u_\|^\star\neq 0,u_\perp^\star\neq 0,
u_\times^\star=0$; (iii) biconical FP ${\mathcal B}$ with
$u_\|^\star\neq 0,u_\perp^\star\neq 0, u_\times^\star \neq 0$.
Below, we will analyze peculiarities of the dynamical critical
behavior in the above  universality classes.

\subsection{Dynamics at the isotropic Heisenberg fixed point}

At the isotropic Heisenberg FP ${\mathcal H}(n_\perp+n_\|)$ the
fourth order static couplings are equal,
$u_\|^\star=u_\perp^\star=u_\times^\star=u^\star$. In consequence
the static couplings drop out in the FP equation for $v$. Assuming a
nonzero finite value of $v$ at the FP, the equation for
$v^\star$, $\beta_v(v^\star)=0$, reads
\begin{eqnarray} \label{heisenbergwstern}
0&=&\Bigg\{\Big[(n_\|+2)-(n_\perp+2)\Big](6\ln\frac{4}{3}-1) \nonumber \\
&-&n_\|\left[\frac{4}{v^\star}\ln\frac{2(1+v^\star)}{2+v^\star}
+2\ln\frac{(1+v^\star)^2}{v^\star(2+v^\star)}-1\right]   \\
&+&n_\perp\left[4v^\star\ln\frac{2(1+v^\star)}{1+2v^\star}
+2\ln\frac{(1+v^\star)^2}{1+2v^\star}-1\right]\Bigg\} \, . \nonumber
\end{eqnarray}
One immediately sees that for general $n_\|$ and $n_\perp$ a zero
can only be found if the arguments of the logarithms are equal to
$4/3$, which leads to  the FP value
\begin{equation}\label{vfix}
v^\star=1 \, .
\end{equation}
This result has to be fulfilled also in higher loop order and therefore the result is exact.
Due to Eq. (\ref{vfix})
the $\zeta$-functions (\ref{zetagperp}) and (\ref{zetagpara}) become equal at the FP
\begin{equation}
\zeta_{\Gamma_\perp}^\star=\zeta_{\Gamma_\|}^\star=(n+2)u^\star(6\ln\frac{4}{3}-1)
\end{equation}
with $n=n_\|+n_\perp$. This means strong dynamic scaling  with the dynamical critical exponent
\begin{equation}
z=2+c\eta
\end{equation}
of the $O(n)$-symmetric model A universality class, where
$c=6\ln\frac{4}{3}-1$ in two loop order,{\cite{dyn}} and  $\eta$ is
the anomalous dimension of the  $O(n)$-symmetric model.

\subsection{Dynamics at the decoupling fixed point}

At the decoupling FP ${\mathcal D}$ static critical behavior does
not fulfill scaling due to the existence of two different
correlation lengths, each for one of the decoupled parts  of the
system (the parallel and perpendicular components of the OP). The FP
value of the static coupling $u_\times$ is equal to zero. Therefore
the flow equation (\ref{flow}) at the static FP reduces to
\begin{equation}
l\frac{d v}{d l}=\frac{v}{72}\Big([(n_\|+2)u_\|^{\star 2}-(n_\perp+2)u_\perp^{\star 2}](6\ln\frac{4}{3}-1)\Big)
\end{equation}
leading either to  a flow reaching $v^\star=0$ or $1/v^\star=0$ depending wether
\begin{eqnarray}  \label{inequ}
n_\|<n_\perp \qquad &\mbox{where}& \qquad   \zeta_{\Gamma_\|}^{\star(A)}<\zeta_{\Gamma_\perp}^{\star(A)}  \qquad\mbox{or}
\nonumber \\
n_\|>n_\perp \qquad &\mbox{where}& \qquad
\zeta_{\Gamma_\|}^{\star(A)}>\zeta_{\Gamma_\perp}^{\star(A)}  \, ,
\end{eqnarray}
with $\zeta_{\Gamma_{\|,\perp}}^{\star(A)}$ the $\zeta$-function of model A for the corresponding subsystem with
$n_{\|,\perp}$ components.
Both cases mean that weak scaling holds at the decoupling FP. Indeed inserting the values of the decoupling FP
into the $\zeta$-functions (\ref{zetagperp}) and (\ref{zetagpara})
gives two dynamical exponents $z$. One for the dynamics of the
parallel and another for the perpendicular components of the OP.
Both exponents correspond to the model A universality class
\begin{equation}
z_\|=2+c\eta_\| \qquad \mbox{and} \qquad z_\perp=2+c\eta_\perp \, .
\end{equation}
In the special case when $n_\|=n_\perp$ the exponents $z_\|$ and $z_\perp$ become equal and the FP values of
the time scale ratio are determined by the initial values of the static flow.

\subsection{Dynamics at the biconical fixed point}

At the biconical FP ${\mathcal B}$ the FP values of the three fourth
order couplings are different and the solution for the FP value of
$v$ becomes nontrivial and dependent on the number of components
$n_\|$ and $n_\perp$. However the only relevant case where this FP
is stable in $d=3$ is $(n_\|,n_\perp)=(1,2)$ or
$(n_\|,n_\perp)=(2,1)$.\cite{note1} Due to symmetry the two
solutions are related
\begin{equation}
v^\star(n_\|,n_\perp)=\frac{1}{v^\star(n_\perp,n_\|)} \, .
\end{equation}
The numerical solution found using the static one loop order FP values for the static couplings reads \cite{dohmjanssen77}
\begin{equation}
v^\star(1,2)=v^{{\cal B}}=1.0241.
\end{equation}
Thus one finds strong scaling with a new biconical dynamical exponent. Inserting the FP values into the
$\zeta$-functions (\ref{zetagperp}) and (\ref{zetagpara}) the dynamical critical exponent is given by
\begin{equation}
z^{{\cal B}}=2.0149.
\end{equation}
As has been shown in part I the biconical FP becomes stable in two
loop order within a small region around the OP-component values
$n_\perp=2$ for $n_\|=1$.
\begin{figure}[h,t,b]
      \centering{
       \epsfig{file=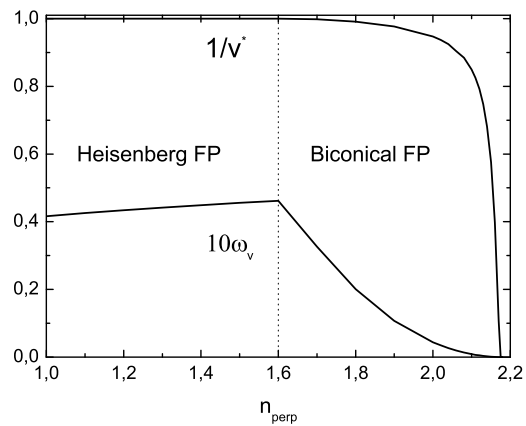,width=8cm,angle=0} }
     \caption{ \label{dynFP} Dependence of FP values of the timescale ratio  $1/v^\star$ and the dynamic transient
exponent $\omega_v$ on $n_\perp$ for $n_\|=1$ in the region of stability of the Heisenberg ${\cal H}(3)$ and biconical ${\cal
B}$ FP in
$d=3$. The dotted vertical line indicates the stability border between ${\cal H}(3)$ and ${\cal B}$. At the stability border
to the decoupled FP both $1/v^\star$ and $\omega_v$ go to zero (at $n_\perp\sim 2.18$). }
\end{figure}
However one has to apply resummation techniques to the two loop
functions in order to get real FP values    for the static
couplings. Using these {\it two loop order resummed} values the
dependence of the FP value of the time scale ratio $v$ within this
region is shown in Fig. \ref{dynFP}. Note that we do not resume
the expression for $\beta_v$, Eq. (\ref{betaw}), itself.  The
biconical FP reaches the value of the Heisenberg FP, $v^{{\cal
H}(3)}=1$, at the stability border line and the decoupled FP value,
$v^\star=0$ or $v^\star=\infty$ (depending on wether $n_\|$ is
larger or smaller than $n_\perp$) at the corresponding stability
border line. Inserting for $n_\|=1$ and $n_\perp=2$  the resummed FP
values for static fourth order couplings into Eq. (\ref{betaw}) one
obtains
\begin{equation} \label{biconvstar}
v^\star=v^{{\cal B}}=1.0555.
\end{equation}
The corresponding dynamical critical exponent reads
\begin{equation}
z^{{\cal B}}=2.052.
\end{equation}
This has to be compared with the predicted value  for the  Heisenberg FP \cite{dohmjanssen77} $z^{{\cal H}(3)}=2.015$,
which was found to be stable in the $\epsilon$-expansion in one loop order .

\section{Dynamic transient exponents \label{dynstab}}

The static stability boundaries are also dynamic stability
boundaries and therefore in the  case $n_\|=1$ and $n_\perp=2$ a
small dynamic transient exponent is expected. Its value is given by
\begin{eqnarray} \label{dyntrans}
\omega_v&=&\left(\frac{\partial \beta_v}{\partial v}\right)_{u_\times^\star,v^\star} \\
&=&u_\times^{\star 2}\frac{v^\star}{18}\bigg(\frac{n_\|}{v^\star}
\ln\frac{2(1+v^\star)}{2+v^\star}+n_\perp\ln\frac{2(1+v^\star)}{1+2v^\star}\bigg)
\, .   \nonumber
\end{eqnarray}
It will be further numerically evaluated by inserting the Borel
resummed values for static couplings. As will be shown below this
exponent goes to zero only when the dynamical FP changes
from the strong dynamic scaling to the weak dynamic scaling FP. This
is the case when in addition to the change of the stability of
the static  FP, the stability of the dynamic FP is changed.

The instability of the weak scaling FP $v^\star=0$ or $1/v^\star=0$ is defined by a negative dynamic transient exponent
\begin{eqnarray}
\omega_v&=&\left(\frac{\partial \beta_v}{\partial v}\right)_{v^\star=0} \nonumber \\
&=&\frac{1}{72}\Bigg\{\Big[(n_\|+2)u_\|^{\star 2}-(n_\perp+2)u_\perp^{\star 2}\Big](6\ln\frac{4}{3}-1)   \nonumber \\
&-&n_\|\ u_\times^{\star 2}\left[3-2\lim_{v\to
0}\ln(2v)\right]-n_\perp\ u_\times^{\star 2}\Bigg\} \, .
\end{eqnarray}
This shows that in any case where  $u_\times^{\star 2}$  is
different from zero the weak scaling FP is never stable. Thus only
at the decoupled fixed point the dynamic transient exponent might be
positive, if $n_\|>n_\perp$, see Eq. (\ref{inequ}). The transient
exponent then reads
\begin{equation}
\omega_v=\frac{c}{18}\left(\eta_\|-\eta_\perp\right) \, .
\end{equation}

\begin{figure}[h,t,b]
      \centering{
       \epsfig{file=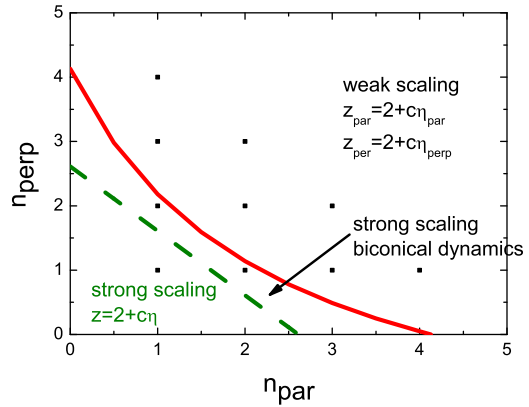,width=8cm,angle=0} }
     \caption{ \label{phasediag} Regions of different static bicritical behavior,
which are defined by the  stable FP  in the
$n_\|-n_\perp$-plane ($\epsilon=4-d=1$) separated by the static stability border lines,
(from left to right: Heisenberg FP, biconical FP and
decoupled FP). The solid line is also the border line between dynamic strong and weak scaling.
 The dots indicate small integer values of the component numbers.}
\end{figure}

At the Heisenberg FP the dynamic transient exponent reduces to
\begin{equation}
\omega_v^{{\cal H}}=\frac{n (u^{{\cal H}(n)})^2}{18}\ln\frac{4}{3}
\, .
\end{equation}
Thus the {\it dynamic} transient exponent at the stability
borderline to the  biconical FP is {\it finite} and {\it
continous} (see Fig. \ref{dynFP}). In the region of stability of the
biconical FP the dynamic transient exponent is given by the
expression (\ref{dyntrans}) evaluated with the appropriate FP values
for $u_\times$ and $v$. At the stability borderlined to the
decoupling FP both FP values go to zero. Thus {\it also} the dynamic
transient exponent goes to zero indicating the change from the
stability of the strong scaling dynamic FP to the weak scaling FP.
Inserting the FP value for the  biconical FP leads to a dynamic
transient exponent roughly one order smaller than at the Heisenberg
FP due to the smaller FP value of the static coupling $u_\times$.
Inserting the FP values for the biconical FP into (\ref{dyntrans})
one obtains
\begin{equation}  \label{transbicon}
\omega_v^{{\cal B}}=0.0044 \, .
\end{equation}
Thus in addition to the already small transient from statics an even smaller transient in dynamics appears.
This leads to a slow approach of the FP values in the flow equations.

The resulting 'phase diagram' concerning the dynamical universality
classes  is shown in Fig. \ref{phasediag}. A strong dynamic scaling
part at small values for the OP components is separated by a
stability border line (solid curve) at which the dynamical transient
$\omega_v$ goes to zero. This border line lies very near the dot
representing the model describing the critical behavior of a three
dimensional Heisenberg antiferromagnet in a magnetic field
($n_\|=1$, $n_\perp=2$). In consequence the transient from the
background to the asymptotic behavior might be very slow.
\begin{figure}[h,t,b]
      \centering{
       \epsfig{file=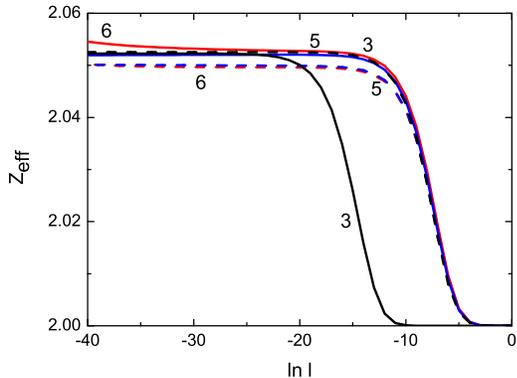,width=8cm,angle=0} }
     \caption{ \label{z1} Effective dynamic exponent for the static flows 3, 5 and 6 (see Fig. 1 in part I). The timescale ratio
$v$ is set to its biconical FP Eq. (\ref{biconvstar}).}
\end{figure}

\section{Flow equations and effective exponents \label{nonasymp}}

The asymptotic dynamic exponents may be reached only in very small region around the FP  where the deviation from the
FP values in the model parameters have died out.
Due to the small transient exponents (either static and/or dynamic) in the physical accessible region the critical behavior
may be an effective one described by effective exponents calculated with the parameters different from their FP values at
$l=0$ and obtained from the flow equations at finite values of $l$.
\begin{figure}[h,t,b]
      \centering{
       \epsfig{file=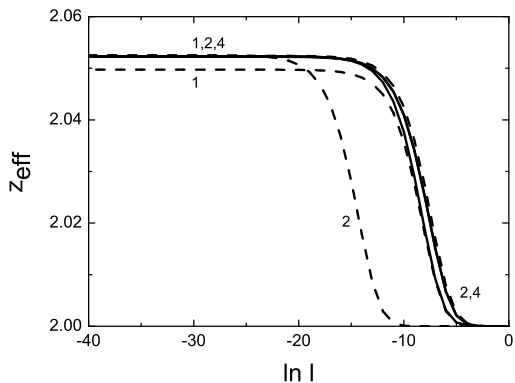,width=8cm,angle=0} }
     \caption{ \label{z2} Effective dynamic exponent for the static flows 1, 2 and 4 (see Fig. 1 in part I). The timescale ratio
$v$ is set to its biconical FP Eq. (\ref{biconvstar}).}
\end{figure}
The effective exponents are defined as
\begin{eqnarray}
z_{eff, \perp}(l)=2+\zeta_{\Gamma_\perp}(u_\perp(l),u_\|(l),u_\times(l),v(l))  \, ,
\end{eqnarray}
\begin{eqnarray}
z_{eff,\|}(l)&=&2+\zeta_{\Gamma_\|}(u_\perp(l),u_\|(l),u_\times(l),v(l))
\, .
\end{eqnarray}
In Figs. \ref{z1} and \ref{z2} we show the effective dynamic
exponents for the parallel and perpendicular components of the OP.
In the asymptotics when reaching the stable biconical FP both
exponents reach the same value since the strong scaling dynamic FP
is stable. In order to show the effect of the small static transient
exponent we fix the value of the time scale ratio  to its biconical
FP value. The initial values for the static couplings are chosen to
be the same as in part I for the flows in Fig. 1 numbered from 1 to
6. Although the static FP value is not reached (see Fig. 1 in part
I) the numerical differences in the effective dynamical exponents
are small. The difference between the parallel and perpendicular
effective dynamical exponent for curves number 2 and 3 in the
background region (larger $l$) result from that part of the static
flow where $u_\times$ and either $u_\|$ or $u_\perp$ are almost
zero.

In order to show the effect of the smaller {\it dynamical transient exponent} (\ref{transbicon}) we fix the static couplings to
their biconical FP
values and start the flow for the time scale ratio $v$ at three different initial values corresponding to situation where the
parallel relaxation coefficient is smaller, equal or larger to one.
\begin{figure}[h,t,b]
      \centering{
       \epsfig{file=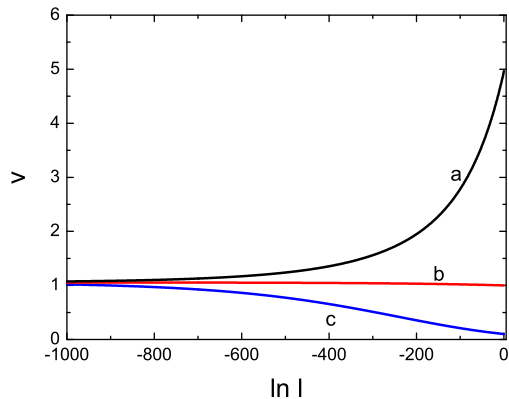,width=8cm,angle=0}
       }
     \caption{Flows of the timescale ratio $v(l)$ at the biconical static FP ${\cal B}$ for different initial conditions.
      $v(l=1)=0.1$,  $v(l=1)=1$ and $v(l=1)=5$. Note the scale of the flow parameter compared to Figs.
\ref{z1},\ref{z2},\ref{z3}.\label{flowv}}
\end{figure}

\begin{figure}[h,t,b]
      \centering{
       \epsfig{file=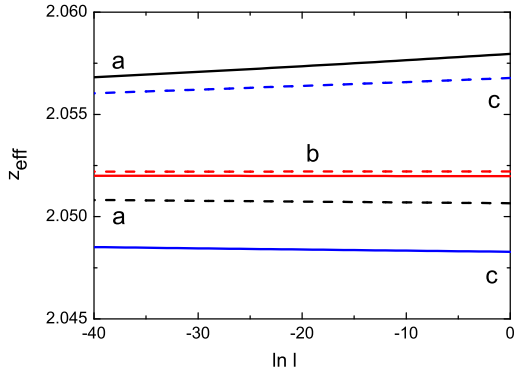,width=8cm,angle=0} }
     \caption{ \label{z3} Effective dynamical exponents $z_{eff, \|}$ (solid curves) and $z_{eff, \perp}$ (dashed curves) for the flows shown in Fig. \ref{flowv}.}
\end{figure}

As one can see from the Fig. \ref{flowv} indeed for  initial values of
$v(0)$  far from its FP value, the timescale ratio almost never attains its FP value (for
$v(1)=0.1$ it reaches the asymptotics at $\ln \ell \simeq -10^3$)! However the  effective exponents $z_{eff}$ are not so far
from their FP values in consequence of the general dependence on the timescale ratio (see Fig. \ref{z3}). However one might define a dynamic amplitude
ratio from the relaxation rates $\Gamma_\|/\Gamma_\perp$. This ratio then would in leading order behave like $v$.

\begin{figure}[h,t,b]
      \centering{
       \epsfig{file=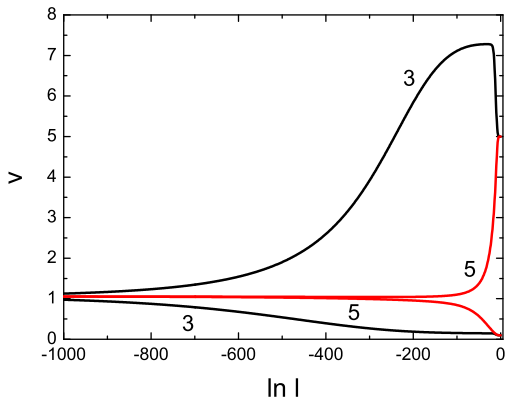,width=8cm,angle=0}
       }
     \caption{Flows of the timescale ratio $v(l)$ to the biconical static FP ${\cal B}$ for different initial conditions.
     3: $v(l=1)=0.1$, 5: $v(l=1)=5$. A quite different behavior is observed. \label{flowv2}}
\end{figure}

\begin{figure}[h,t,b]
      \centering{
       \epsfig{file=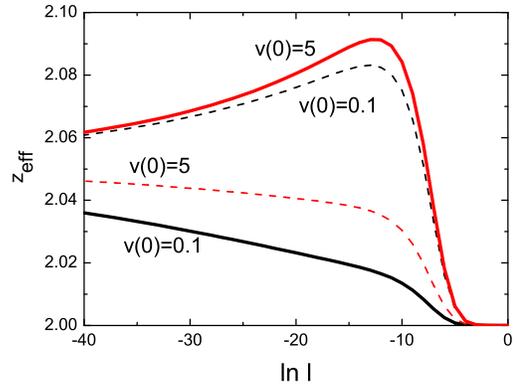,width=8cm,angle=0}
       }
     \caption{Effective dynamic exponent $z_{eff,\|}$ (solid curves) and $z_{eff,\perp}$ (dashed curves) for the
static initial conditions of flow number 5 and dynamic initial conditions as indicated. For the different behavior of the
parallel and perpendicular exponent see the text. \label{zflow5}}
\end{figure}

Starting the flows of 3 and 5 with $v$ different from its FP value (see Fig. \ref{flowv2}) leads to nonmonotonic behavior
of the effective dynamic exponents (see Fig. \ref{zflow5}).
At the first sight the behavior of the parallel and perpendicular
effective dynamical exponents looks strange since the nonmonotonic
behavior is seen in Fig. \ref{zflow5} for small initial values
$v(0)$ in the perpendicular exponent, whereas for large values
$v(0)$ in the parallel exponent. To explain such an unexpected
behavior, one may look at the difference of the two effective
exponents (see Eqs (\ref{zetagperp} and \ref{zetagpara})),
 \begin{equation}\label{1}
  \Delta z = z_{eff,\|}(u_\|, u_\times, u_\perp, v) - z_{eff,\perp}(u_\|, u_\times, u_\perp, v)
 \end{equation}
 and estimate it at fixed $u_\|, u_\times, u_\perp$ but for
 different values of $v$. The difference can be written as:
\begin{equation} \label{2}
\Delta z = c + \delta z(v) .
\end{equation}
In Eq. (\ref{2}), $c$ depends on the static coupling only and
$\delta z(v)$ for $n_\perp=2$, $n_\|=1$ reads:
\begin{eqnarray} \label{3}
\delta z(v)&=&
 \frac{{{\it u_\times}}^{2}}{36} \left( 4{{\it v}}\,\ln  \left( 2\,{\frac {1+{\it
 v}}{1+ 2
{\it v}}} \right) +2 \ln  \left( {\frac { \left( 1+{\it v} \right)
^{2}}{ 1+2{\it v} }} \right) \right. \nonumber \\
&-& \left. \frac{2}{{{\it v}}}\,\ln \left( 2\,{\frac {1+{ \it
v}}{2+{\it v}}} \right) -\ln  \left( {\frac {
 \left( 1+{\it v} \right) ^{2}}{{\it v}\, \left( 2+{\it v} \right)
}} \right) \right)
\end{eqnarray}
depending on $v$ and the static coupling $u_\times$. For flows where
$u_\times$ is very small no difference is seen according to $v$.
So no difference is seen in the corresponding flow for the static flow number 3.

However for $u_\times$ near the bicritical FP,  the difference between
$z_\|$ and $z_\perp$ calculated at the {\em same values of the
static couplings} but at {\em different values of the timescale
ratio} sometimes can be positive (i.e. $z_\|> z_\perp$) and
sometimes it can be negative ($z_\| < z_\perp$), depending on
particular values of $v$. Numerically estimates indeed recover the differences
shown in Fig. \ref{zflow5}

\section{Conclusion and outlook}
We have reconsidered the relaxational dynamics at the multicritical dynamical FPs in  $O(n_\|)\oplus O(n_\perp)$
symmetric systems. According to the static two loop order results the biconical FP is the stable FP at the interesting case
$n_\|=1$ and $n_\perp=2$ for which a new dynamic FP with strong dynamic scaling has been found.

The critical dynamics of such a system has to take into account additional properties, namely if the densities
of conserved quantities couple statically to the OP and/or if mode coupling term are present.  Both extensions of the
dynamical equations have to be considered lacking a complete two loop order calculation\cite{dohmjanssen77}.  The model C
like extension of this model will be presented in a third part of this series.

Acknowledgement: This work was supported by the Fonds zur F\"orderung der
wissenschaftlichen Forschung under Project No. P19583-N20.


\begin{thebibliography}{99}
\bibitem{note1} In the minimal subtraction RG scheme at $d=3$, this
result is obtained already in the two-loop approximation, if one
properly takes into account the asymptotic propertises of the series
obtained.\cite{partI} It is further  supported by the five-loop
calculations.\cite{capevi03} Noteworthy, a similar analysis of the
RG expansions in the massive scheme shows that the two-loop
approximation is still insufficient: the biconical FP appears to be
stable for $n_\|=1$, $n_\perp=3$ and $n_\|=n_\perp=2$ as
well!\cite{Prudnikov98} The closeness of the stability
borderline between the decoupled FP and the biconical FP to the biconical FP found in
Ref.\cite{partI} is of importance for the effective dynamical
behavior as shown in this paper.
\end{thebibliography}
\end{document}